# Innovative Method for enhancing Key generation and management in the AES-algorithm

**Omer K. Jasim Mohammad**
Al-Ma'arif University College, Ramadi, +964, Iraq
Email: omer.k.jasim@ieee.org

**Safia Abbas, El-Sayed M. El-Horbaty and Abdel-Badeeh M. Salem**

Ain Shams University, Faculty of computer and information sciences, Cairo, +2, Egypt
Email: { safia_abbas, Shorbaty , absalem }@ cis.asu.edu.eg

***Abstract*—**With the extraordinary maturity of data exchange in network environments and increasing the attackers capabilities, information security has become the most important process for data storage and communication. In order to provide such information security the confidentiality, data integrity, and data origin authentication must be verified based on cryptographic encryption algorithms. This paper presents a development of the advanced encryption standard (AES) algorithm, which is considered as the most eminent symmetric encryption algorithm. The development focuses on the generation of the integration between the developed AES based S-Boxes, and the specific selected secret key generated from the quantum key distribution.

***Index Terms*—** advanced encryption standard, QK quantum key distribution, Cryptography, cryptanalysis, pseudo random number, secret key.

## I. Introduction

With the development of electronic and optical fiber communication networks, the quantity of information exchanged and the reliance of organizations on these new communication channels has increased dynamically[1][2]. Concurrently, the risks are increased significantly, therefore, many technologies were developed to cope with these threats.

Information security is one of the essential issues in contemporary computer systems and the encryption process is the main issue. Encryption process appeared before the computer system for many years. Caesar and manoa-alphabitc ciphers are a famous example of traditional cipher [3][4]and [5].

As shown in Fig.1, highly progress in information security is linked to computer systems. Nowadays, modern secure communications generally rely on one of two basic cryptographic techniques to ensure the confidentiality and integrity of traffic carried across the network, symmetric and asymmetric mechanisms. In the symmetric cipher a single key for both processes (encryption/decryption), whereas, two keys instead of one key are used for encryption/decryption process in asymmetric cipher [6, 7].

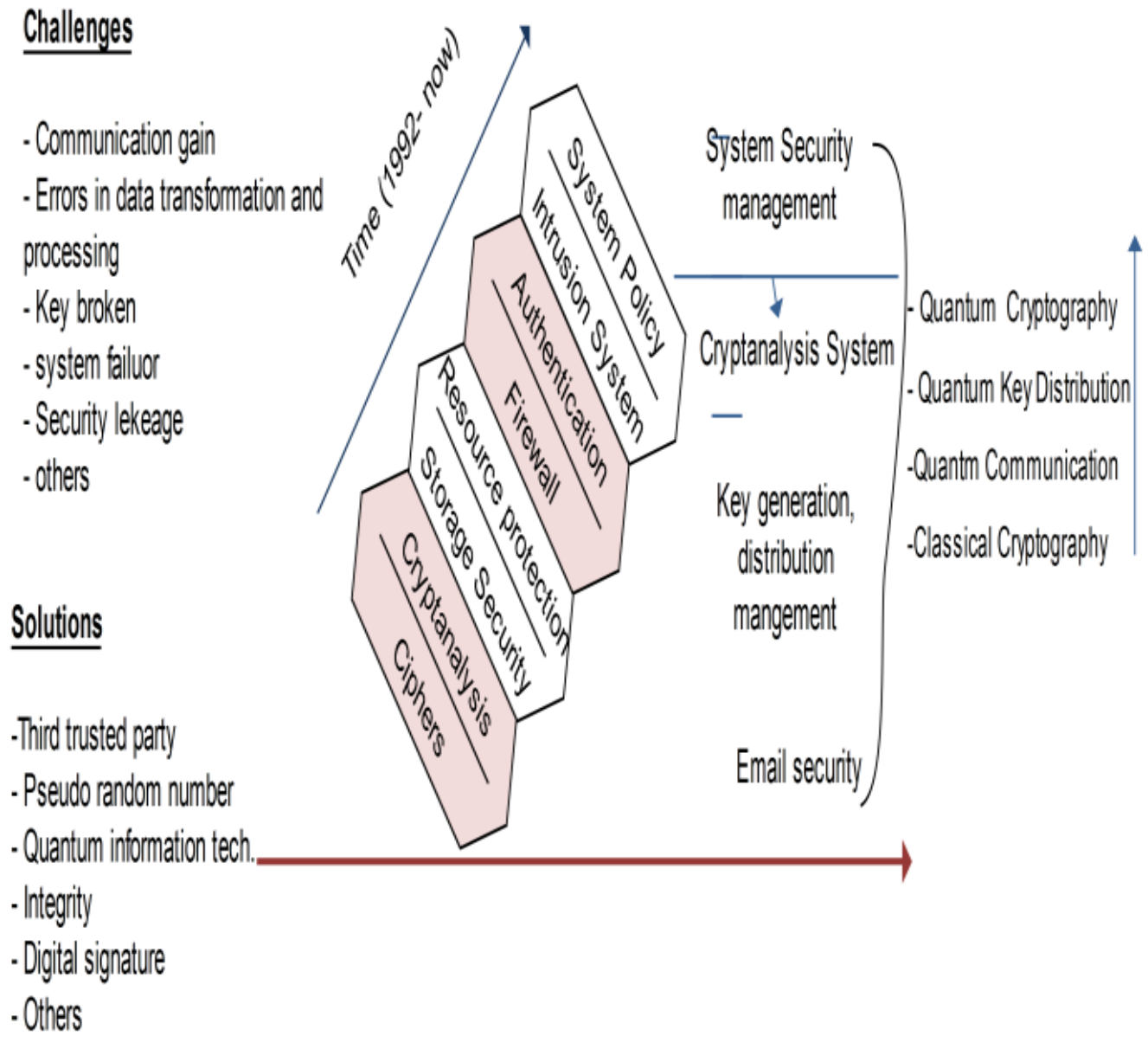

Fig. 1. Security progress maturity

Consequently, symmetric ciphers are divided into two broad categories stream ciphers and block cipher. In a stream cipher, an encryption /decryption process embraces on a symbol by symbol at a time. However, block cipher groups the plain text as a block (B>1) and encrypting together and recur at a decryption process [8, 9].

Generally, in the symmetric cipher the information is sent as cipher text using an unsecured channel, and a secure channel is used to send the key. However, the secure transfer of encryption keys remains one of the critical problems in modern cryptography [10][11].

This paper discuss one type of strong symmetric block ciphers, it is AES, because AES provides strong encryption (NIST selection), high speed algorithm, low memory costs, easy to implement and secure against many analysis attacks such brute-force, differential and linear attacks [9, 12].

AES algorithm utilizes same key for encryption/decryption process, key length is 128; 192; 256- bits embraces static input data block of 128-bits inception as 4*4 matrix, it's known state [13], the length of the key determined the number of rounds (Nr) 10; 12;14. Finally, this paper presents a new version of AES by integration between an enhanced version of AES and QKD [for more details about AES see [9]].

The rest of the paper is organized as follows: Section 2 surveys the existing study for AES optimization, Section 3 presents the details of QKD mechanism. Our experimental setup, QAES architecture, and integration methodology are given in Section 4. Section 5 explains the inferences obtained from the results and analysis it. Section 6 presents the conclusion and future works.

## II. Exising Study

Recently, there are few authors are focused on the enhancing and developing the AES algorithm, because most of AES attackers appeared lately.

For example, Sumalatha et al. [1] design of an Encryptor (128-bits) using AES (128bits) data encryption and 128 bits for a cipher key. Also, the analysis and synthesis processes provided based on VeriloG and Xilinx ISE 14.2 software. This Encryptor bolstered the high secure for ciphering system, however, the fixed S-Box transformation consumes the more memory allocations.

Shaaban et al. [3, 4] develop a powerful algorithm for cryptography, it's based on AES to generate different sub keys from the original key and using each of one to encrypt single AES round. It is resistant against the analysis attacks such as a brute force attack and others. Moreover, the authors classified the secret keys as real key and PRNG, each one used with special cryptographic mode. However, this algorithm is slower than classical AES, therefore, it is apt to the timing attack.

Leonard W.[25] presents scalable system that combines high speed of modern encryption algorithm with the power of quantum key distribution (QKD) technology which annotated as Cerberis. Cerberis is a system that offers a radically new approach to network security based on a fundamental principle of quantum physic. In this study, the author doesn't present the complete simulation environment for Cerberis system.

Kazys et al. [2, 16] present a new version of AES by generating random S-Boxes coinciding with every secret key generation. The authors set out in details how to generate random S-Box, key-independent and then computed the ratio of independency for the S-Box elements. The weakness of this study was not debating any type of cryptanalysis attacks.

Sekar et al. [5] propose a new innovative method to enhance the AES algorithm by increasing the key length to 512 bits and thereby the number of rounds is increased in order to provide a stronger encryption method for secure communication. Code optimization is performed, in order to improve the speed of encryption/decryption using the 512 bit AES. This method don't modify the structure of AES but only increase the number of rounds and so the attacks which need the same key are still serious to this algorithm, at the same time this algorithm increase the processing time which will limit the use of AES in real applications.

Finally, we can conclude that the AES is an efficient modern encryption algorithm. However, it suffers from the key generation, distribution, and management. In order to address these problems, this article presents a new simulator environment which based on integration between QKD and AES-128 algorithm.

## III. Why QKD?

QKD is a major practical application of quantum information, it is based on laws of physics rather than computation complexity of mathematical problems[18][19]. Quantum information is stored as the state of atomic or sub-atomic particles usually called quantum bits (qubits). A qubit is an elementary unit of quantum information and many physical realizations of qubits such as an electron, photon, and quantum dot [20][21].

In networks environment, light is routinely used to exchange information in a form of light pulses, typically containing billions of particles of light, called photons. However, a single photon based in QKD. QKD is based on qubits, these qubits are single photon polarized into one of four states (vertical (V), horizontal (H), left(LD) and right diagonal(RD)) selected from one of two basis (rectilinear and diagonal).

Moreover, it uses the BB84 protocol to achieve the secret key generation and management over two basic channels, classical and quantum channel see Fig.2. Quantum channel is used for transmission of quantum key material by means of photons. The other, classical (public) channel carries all message traffic, including the cryptographic protocols, encrypted user traffic [for more details see [22][23][24].

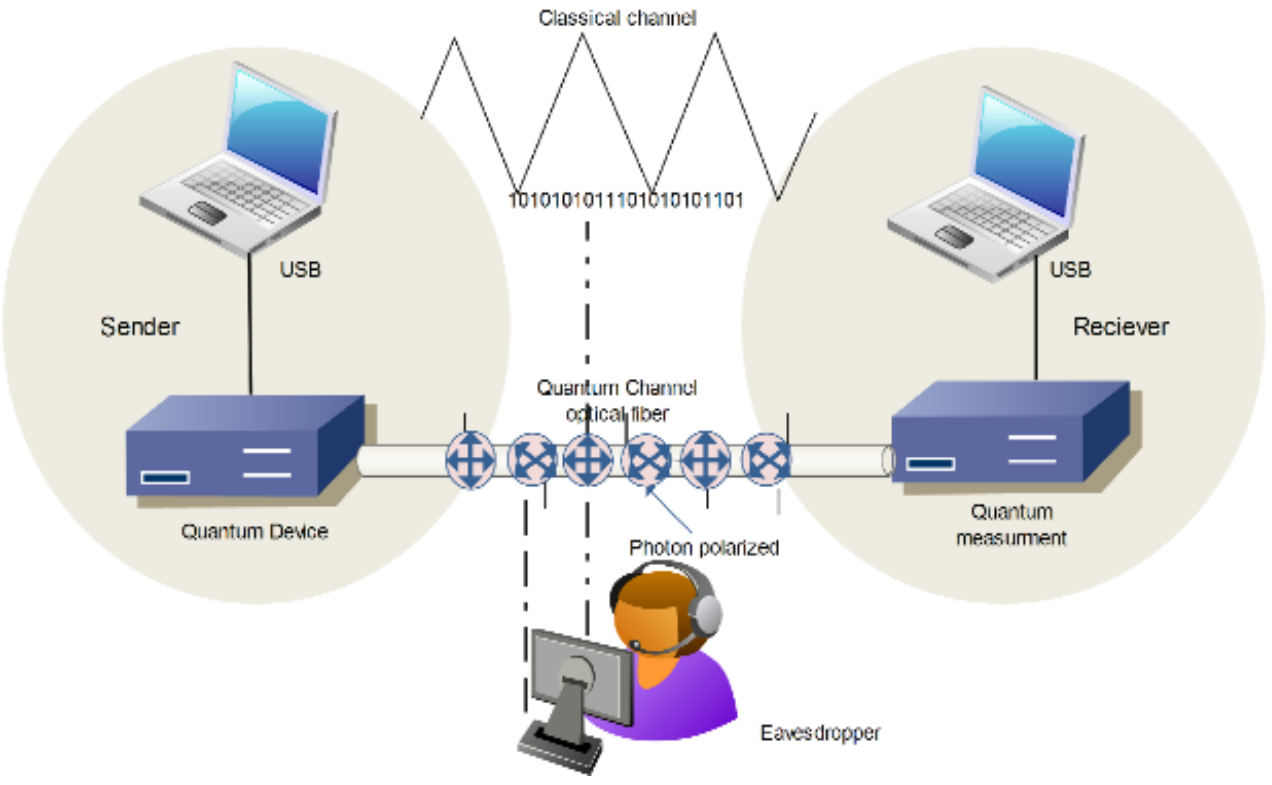

Fig.2. QKD components

Finally, Quantum cryptography relies on the laws of quantum mechanics to provide a secure system, while traditional systems rely on the computational difficulty of the encryption methods used to provide a secure system [25]. However, quantum cryptography is not same QKD.

As shown in Fig. 3, Quantum cryptography is more general and comprehensive, because the QKD is responsible for the key generation and key deployment between two communication parties, while quantum cryptography is the process of encrypting files using one of the usual modern encryption algorithms by using the keys generated from QKD.

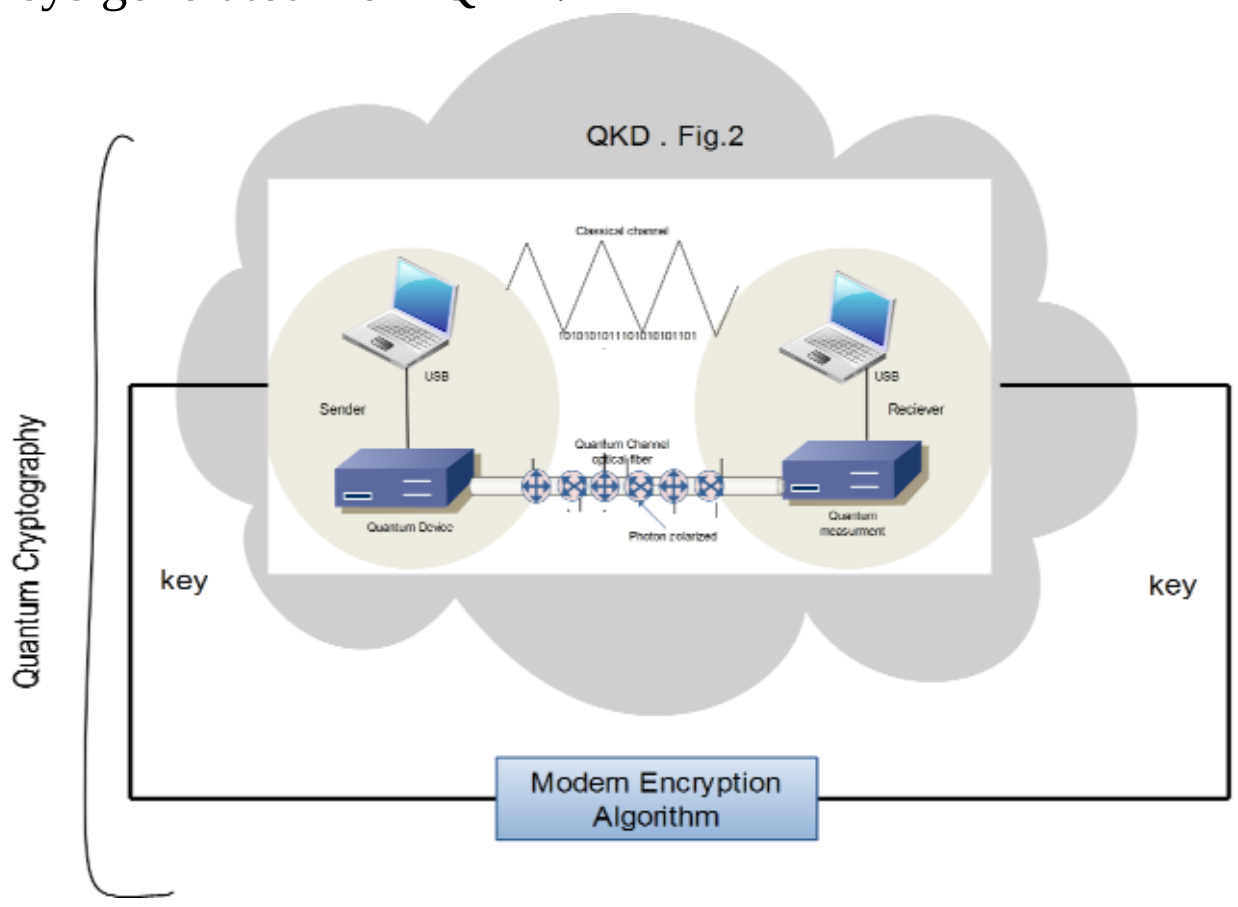

Fig.3. Quantum cryptography architecture

## IV. QAES Architecture

This section illustrates the AES development steps, as shown in Fig. 5, the experimental environment consists of two machines (sender and receiver). The sender machine utilizes the Core i5 (4.8GHz) with 8GB of RAM with 500GB-HDD, while the receiver machine makes use of Core i3 (2.4GHz) with 2GB of RAM with 300 –HDD. The QKD and AES are programmed using Visual Studio Ultimate 2012 (VC#) based Windows Server 2012 Data Center as operating system.

### *A. QAES Single Round*

The QAES developed system incorporates both the QKD and the AES algorithm in order to provide an unconditional security level [25] for any cipher system built on symmetric encryption algorithms. As shown in fig 4, the AES enhanced version exploits the generated key based QKD in the encryption /decryption process. Since the unconditional security depends on the Heisenberg uncertainty principle [19] [20], instead of the complex mathematical model in key generation, more attack resistance is assured and the cipher system is hard to be attacked.

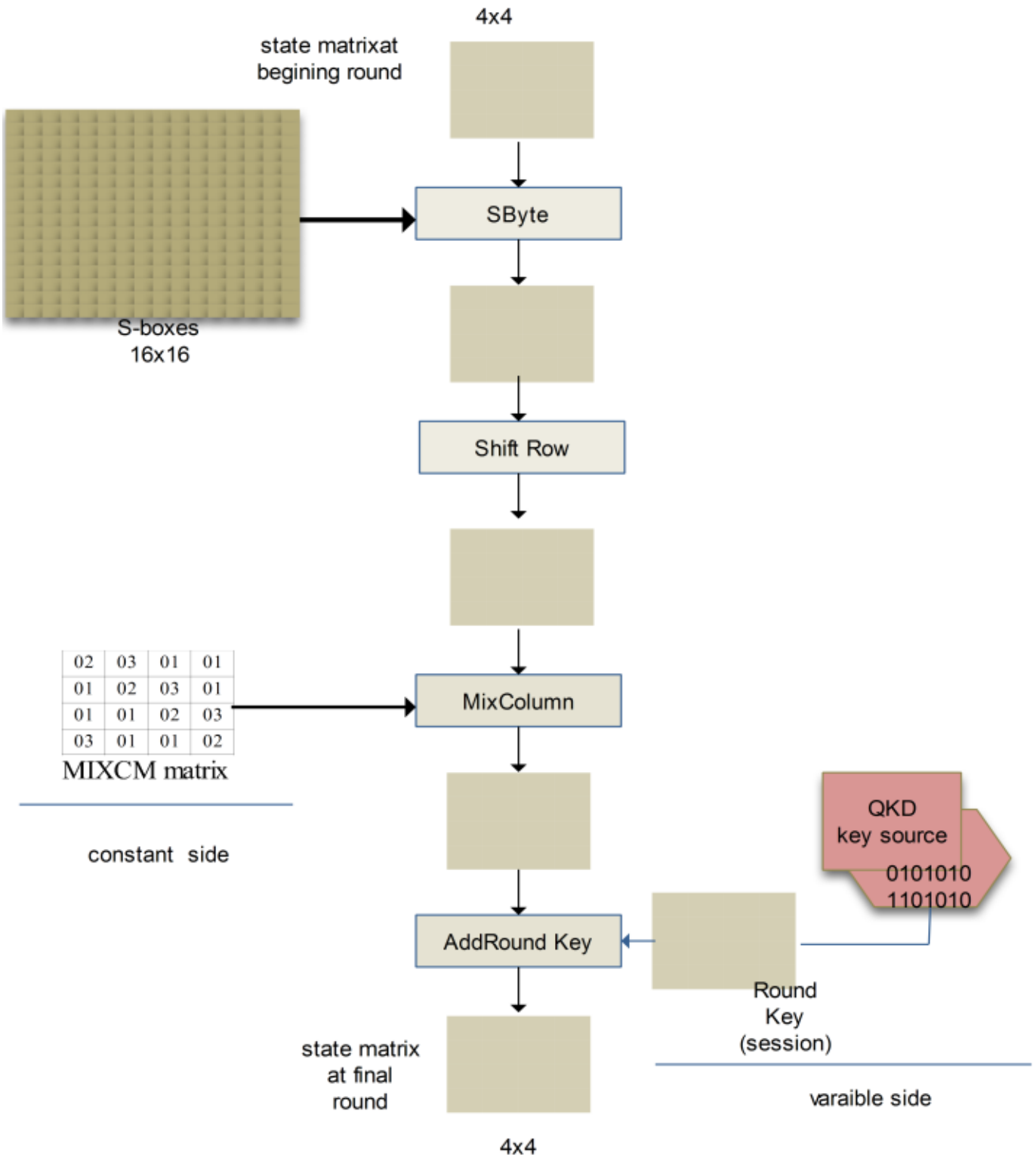

Fig. 4. QAES round

The round key session enjoy the dynamic mechanism, in which the contents of each key session changes consequently in each round with the change of the key generation. Such dynamic mechanism aids in solving the mechanism problems like avoiding the off-line analysis attack, and resistance to the quantum attack.

### *B. Integration Methodology*

This section explains the integration between the enhanced AES and the QKD, during the negotiation between the two parties (sender, receiver). The decryption/ encryption process is achieved coinciding with quantum key generation see Fig.5.

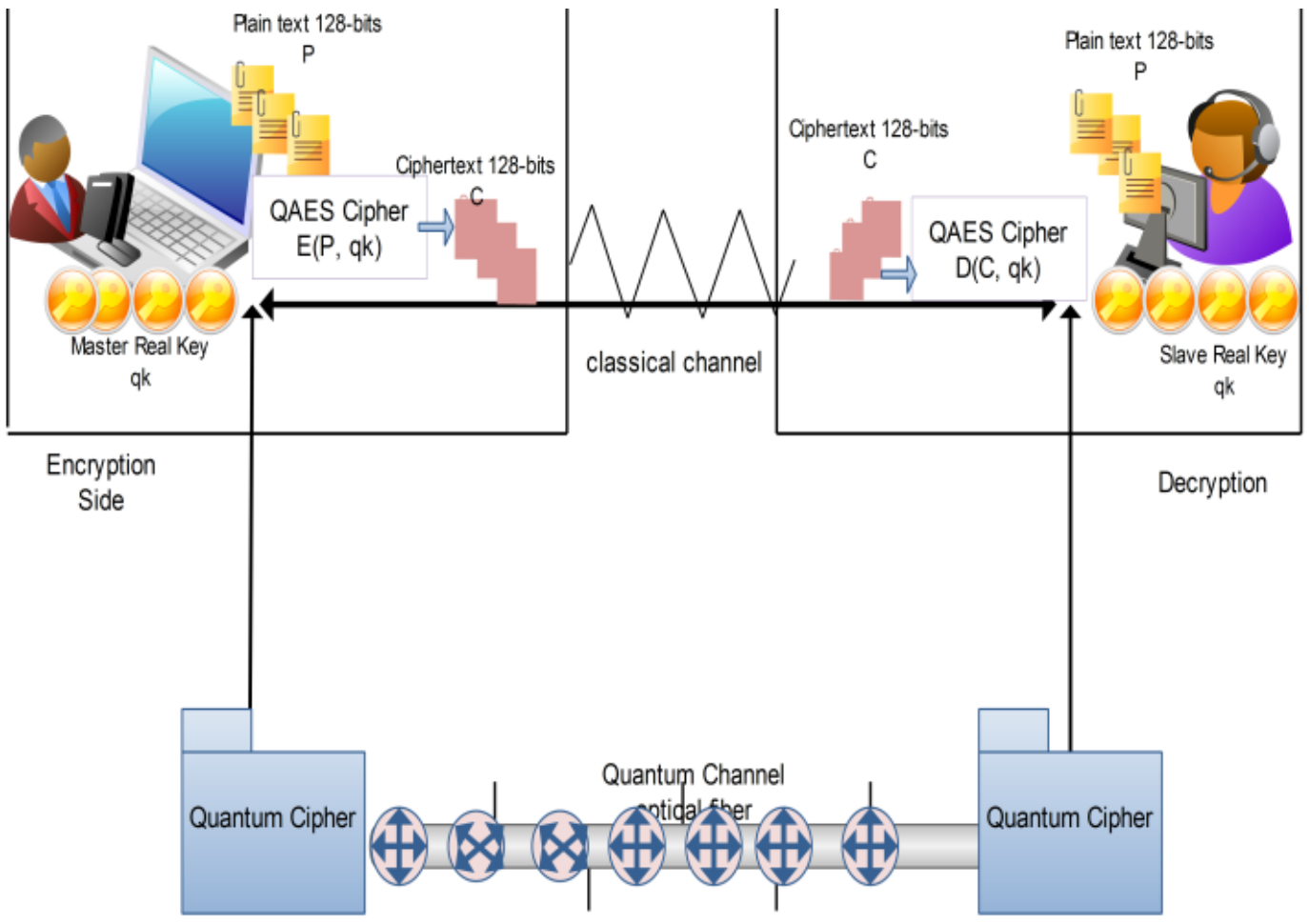

Fig. 5. Integration architecture

*To achieve the encryption /decryption process in QAES-128 must be follow the following steps:*

- The quantum secret key is generated over the quantum channel using BB84 protocol.
- The sender and the receiver parties check the online compatibility for the generated secret key.
- The sender and the receiver choose the appropriate key length (128; 192; 256 bits) through the classical channels in order to perform the encryption/decryption process.
- The two parts deploy the selected final secret key (qk) to the symmetric encryption algorithm (AES).
- Encrypt the first block input file ($P_1$-128bits) by the AES stages- using qk1which generate by QKD round1.

$$E\ (P_1 \oplus qk_1) = C_1$$

- Encrypt the final block input file by the AES stages- using $qk_n$ which generate by QKD $round_n$, where n=Nr=10; 12; 14.

$$E\ (Pn \oplus qkn) = Cn$$

- The decryption process start with the end of the encryption process (inverse methodology).

$$D\ (C_n \oplus qk_n) = P_n$$

Due to the key availability (KA) associated with QKDs[24] , the integration is provided by a sequence of unrelated keys (qk1, qk2,…, qkn) in each round, see Fig. 6, these unrelated keys prevent the attackers from detecting the next key generation. Then, each QAES round, will consider the generated sequence of keys (qk1,qk2,…,qkn) as a sequence of sub keys, which in turn are used in the encryption/decryption process.

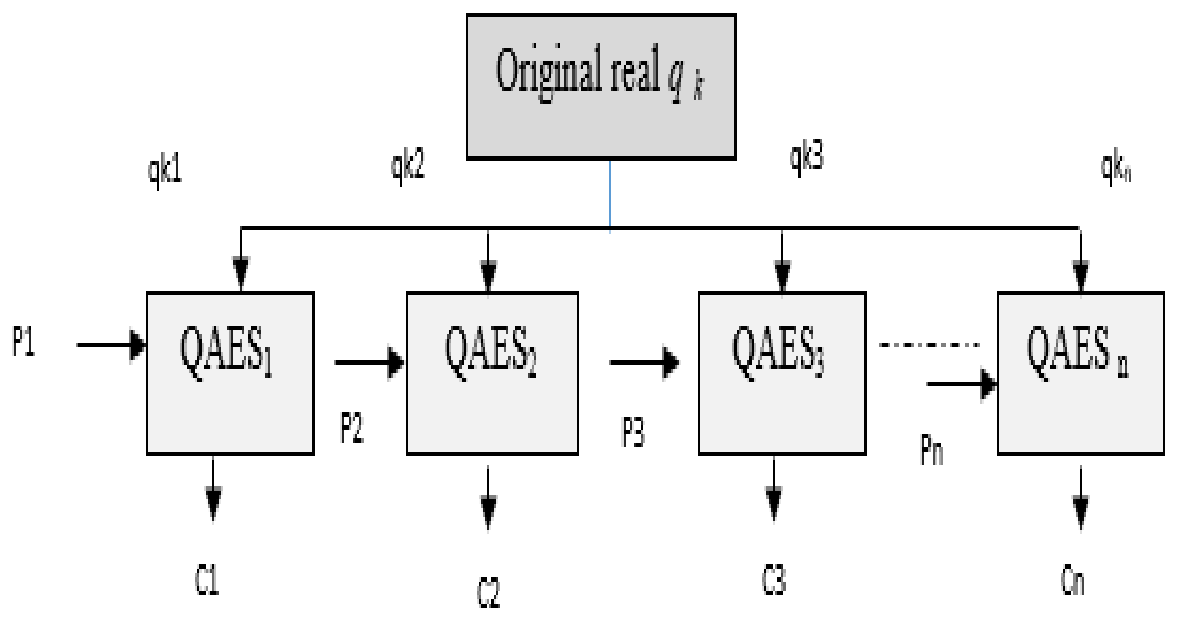

Fig. 6. QAES encryption mode

Finally, this methodology can be used with any type of encryption modes such as cipher feedback (CFB) mode, output feedback (OFB) mode, and counter (CTR) mode[9].

## V. Results and Analysis

In this section, the time of encryption process and NIST testing algorithms has been implemented, measured and analyzed based QAES and AES. The results indicate that the harder for the hacking process and provide a more secured connection.

### A- *Analytical analysis of QAES*

In the following analysis, both the AES and QAES techniques have been implemented using several input files sizes: 500kb, 1000kb, 1500kb, 2000kb, and 3500kb.

#### 1- *Traditional AES (128;192;256) efficiency*

Fig.7 represent the running time of the traditional AES using the local machines described above, the running time is calculated in milliseconds and the input size is taken in kilobytes.

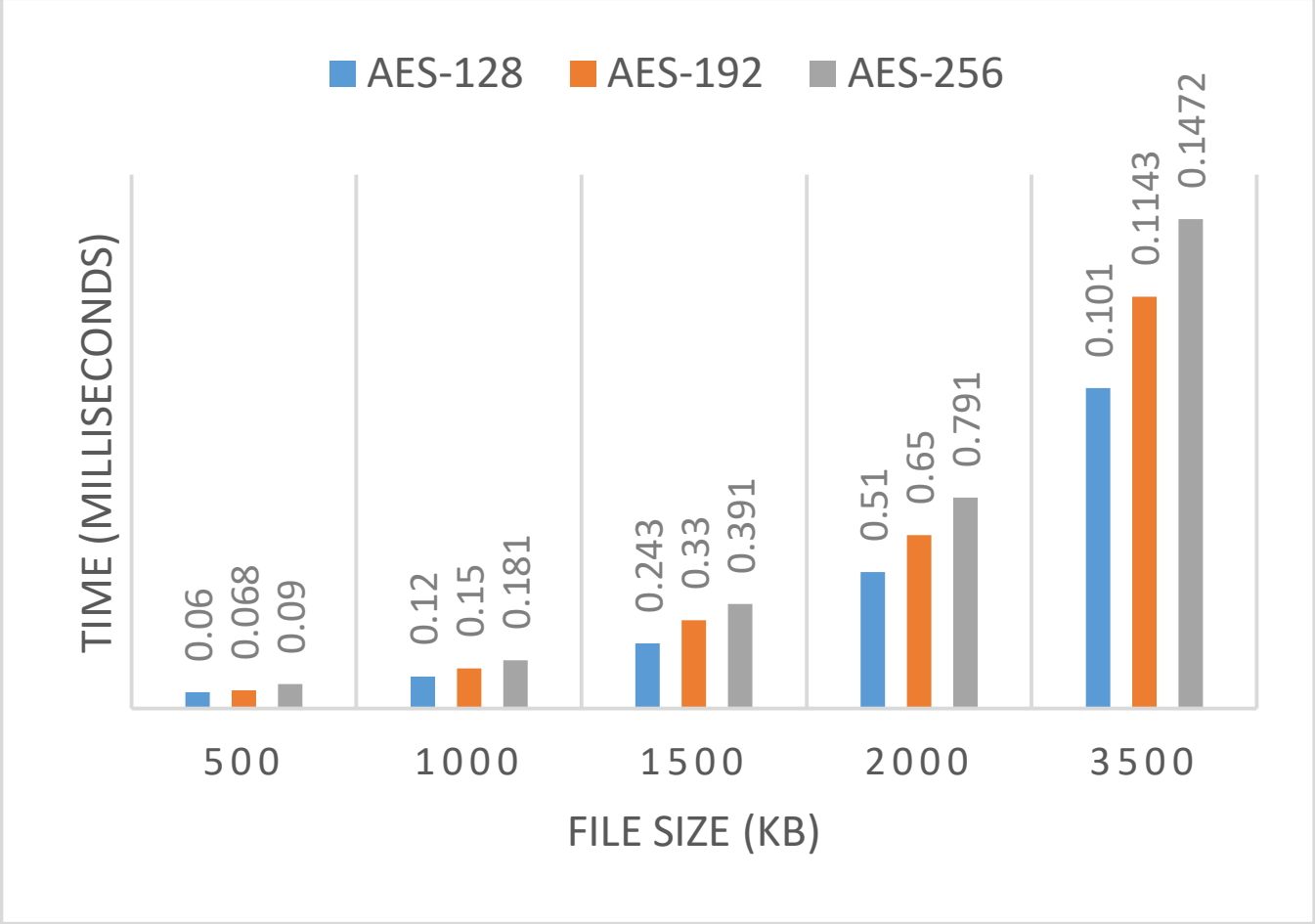

Fig. 7. Running time for classical AES

Based on above figure, there is an inverse proportion relation between the running time and the size of the input file. Such that, the increase of the input file size led to the decrease of the running time. Moreover, the AES is the fastest symmetric technique since it enjoys the scalability based on different hardware, as well as it can be implemented simply. After then, the symmetric techniques can be ordered as DES, 3-DES, RC4, and finally the Blowfish [for more details see [27]].

#### 2- *QAES (128;192;256) efficiency*

Fig. 8 represent the running time of the implemented QAES using the local machines described above, the running time is calculated in milliseconds and the Input size is taken in kilobytes.

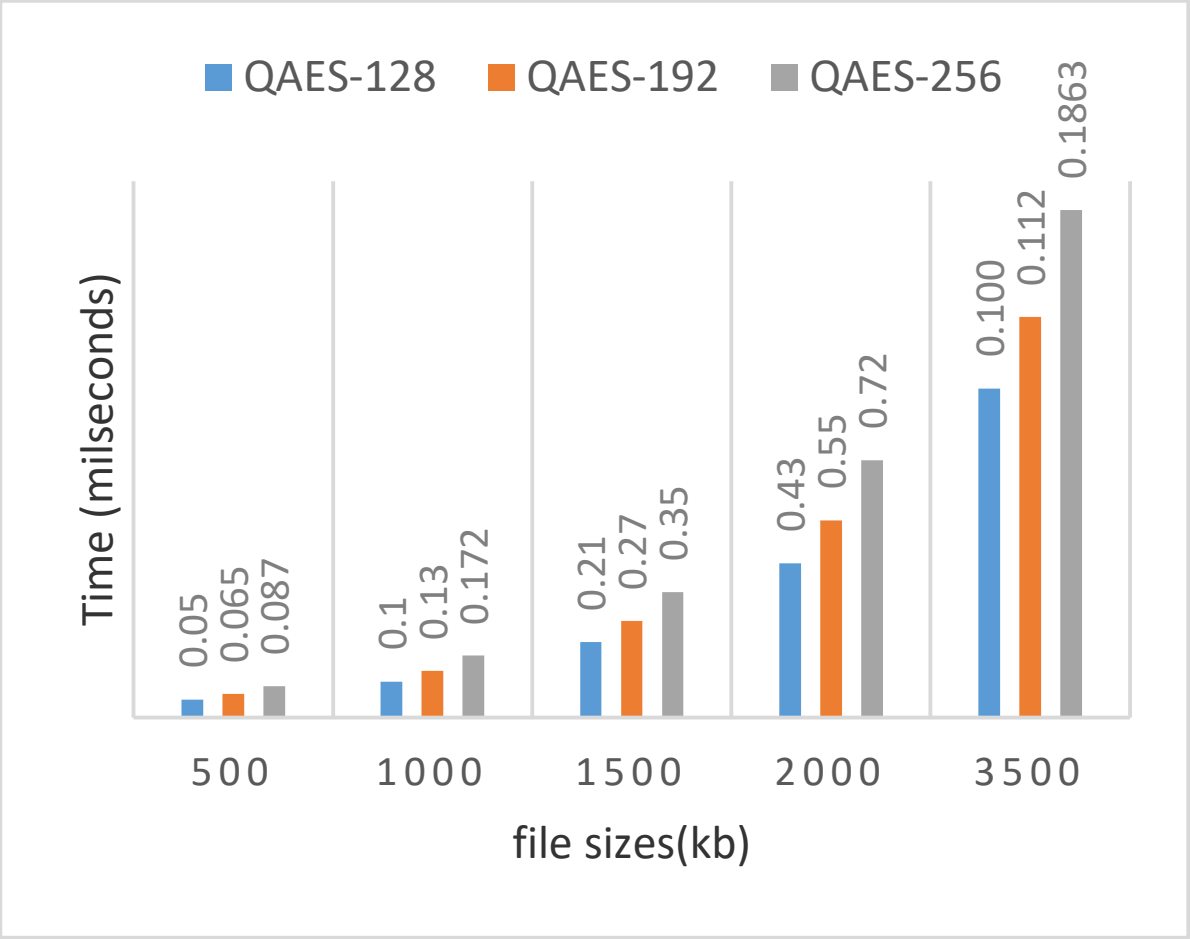

Fig. 8. Running time for QAES without key generation time

Comparing the QAES with traditional AES encryption algorithms reflects a higher security level. However, as shown in Eq.1, and Fig 9, this algorithm takes time more than others due to the time required for quantum key generation (time for quantum negotiation and time required for the encryption / decryption process).

$$T_{qenc} = T_{qkg} + T\ (Enc\ (P)) \qquad (1)$$

*Where* $T_{qenc}$ *= Total encryption based quantum,* $T_{qkg}$ *= time for quantum key generation, T (Enc (P)) = time takes by selecting an encryption algorithm, and P=plain file.*

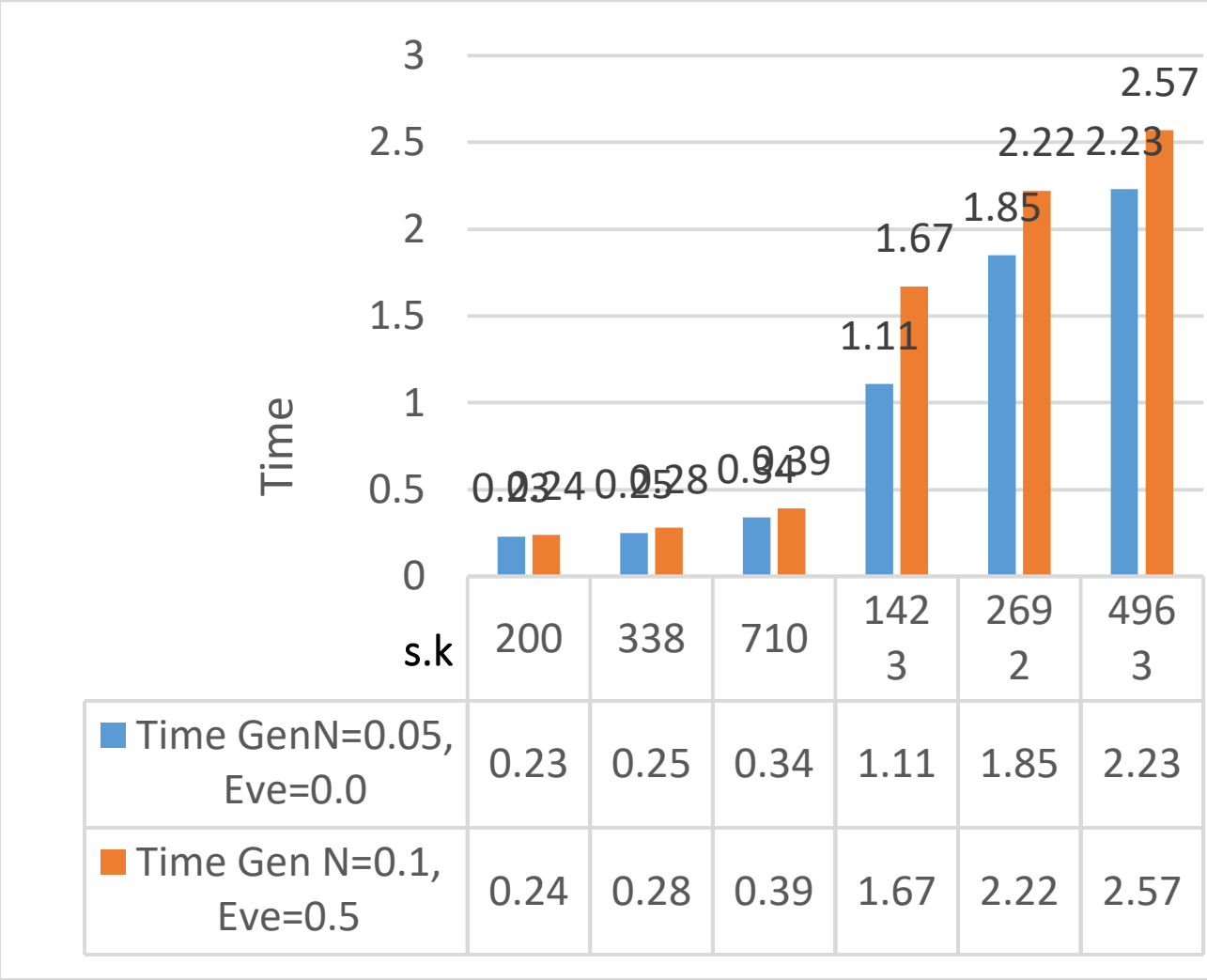

| s.k | 200 | 338 | 710 | 1423 | 2692 | 4963 |
|---|---|---|---|---|---|---|
| Time GenN=0.05, Eve=0.0 | 0.23 | 0.25 | 0.34 | 1.11 | 1.85 | 2.23 |
| Time Gen N=0.1, Eve=0.5 | 0.24 | 0.28 | 0.39 | 1.67 | 2.22 | 2.57 |

Fig. 9. Secret key generation time with various configurations

As showed in Fig. 9, any change in the eavesdropper activity or noise level directly impact on the length of the secret key and the time of generation it. For example, in order to get 200 from 500 qubits are pumped under the impact of noise 0.05 GHz and there is no Eve influence, we need 0.23 milliseconds to generate it. Usually, this time simultaneously grows with the increasing of the noise or Eve influence. However, the practical environment is faster than the simulator environment, due to the light nature [25].

Generally, from the above analysis figures, we conclude that the QAES is a little bit slower than the AES. For example, if we take the sample file 3500 kb, the encryption time for AES is 0.1472 milliseconds, and for QAES is 0.1863 milliseconds (see eq. 1).

Finally, since the QAES follows the same architecture of the AES, the input file size has always changed during encryption process and the details of the processed file remain unchangeable.

### *B- NIST testing*

In this subsection, NIST tests algorithms are implemented to evaluate the security rate of developed algorithm (QAES). NIST developed to test the randomness of a binary sequences produced by either h/w or software based cryptographic random or pseudorandom number generators [for more details see [28]].

Fig. 10 shows the main steps for testing the security of QAES algorithm rely on 12-NIST test algorithms. QKD generates qubits streams as a pseudo random number [28], each sires is (28000,000- bits) and key stream (128-bits) .

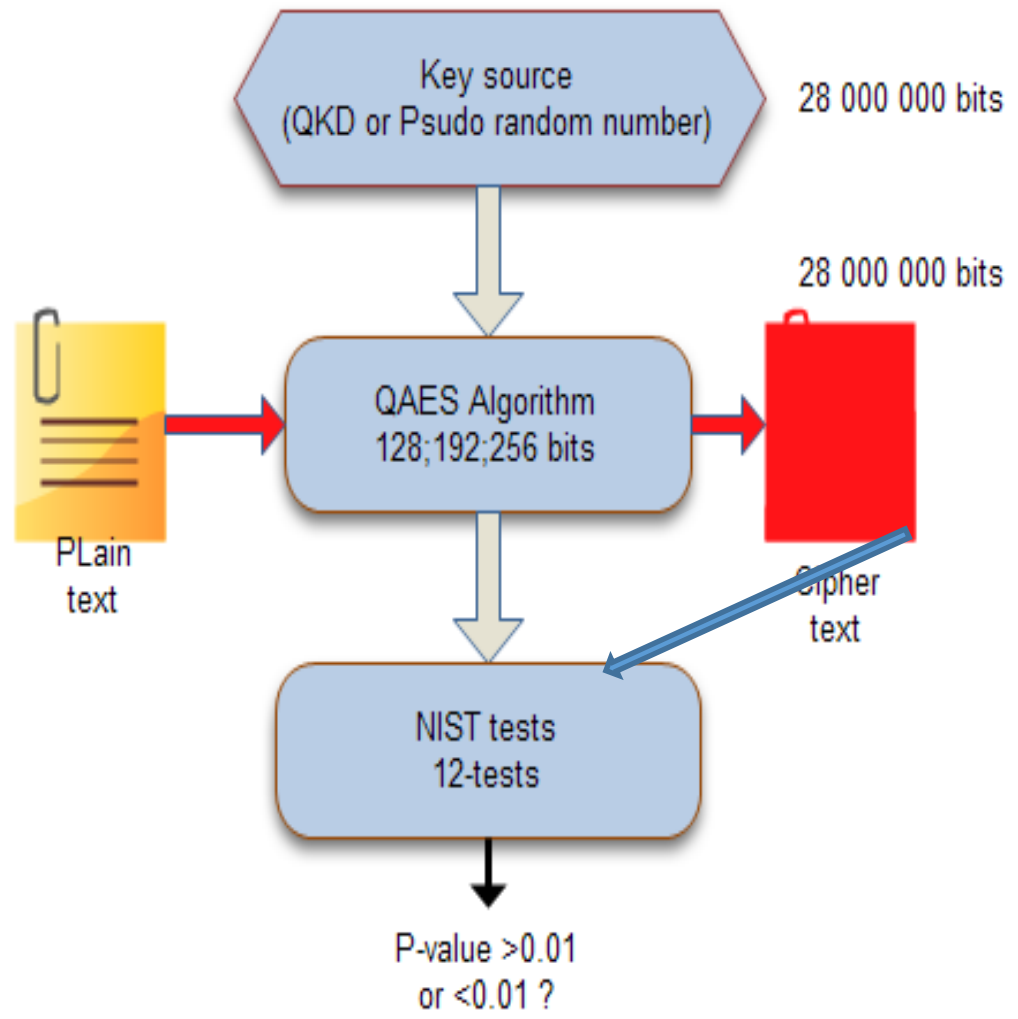

Fig. 10. P-value computation

Now, we applied the QAES to get cipher text and execute NIST tests for each sequence in order to obtain the P-value. After then, compare P-value to 0.01, if p-value less than 0.01 then reject the sequence. The p-value represents the probability of observing the value of the test statistic which is more extreme in the direction of non-randomness, usually, the results are produces in the isolated text file.

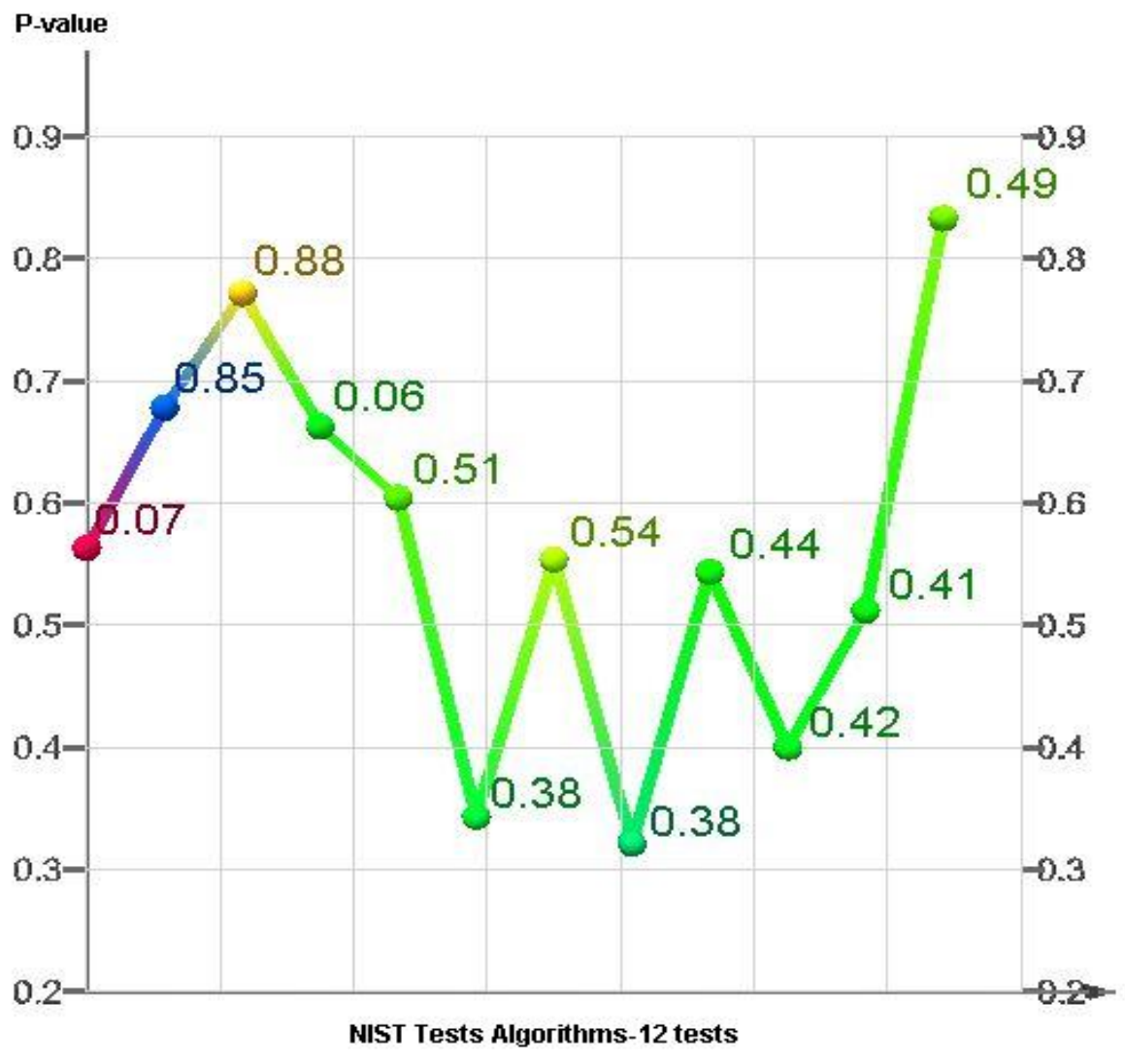

Fig. 11. NIST test for QAES cipher text (12-test)

Regarding to NIST testing algorithms and the P-values are illustrated in fig 11. , we can conclude that the QAE is secure encryption algorithm.

## VI. Conclusion and Future Works

QAES combines strong high-speed encryption (AES-NIST selected) with quantum key distribution and makes it possible to achieve an unprecedented level of security. These techniques are particularly applicable for high value applications and long term secures data retention requirements.

This paper shows that the QAES development and design do not contradict the security of the AES algorithm, since all the mathematical criteria remain unchanged. The QAES symmetric encryption algorithm has been revealed depending on the integration between the AES and the QKD. The experimental results and the analysis show that the QAES produces more complicated un-breakable keys, hard to be predicted by attackers than the keys generated by the AES. However, the speed of encryption of the QAES is tiny slower (0.409 seconds) than using the AES due to quantum key generation.

The strength of the QAES lies in its ability of generating a high ratio of independence between key rounds. Moreover, depending on NIST tests algorithms, QAES achieves the diffusion- confusion principles, this aids in achieving a more secured environment against most types of cryptanalysis attacks.

In the future, firstly the QAES is going to be implemented and tested using cloud environment, secondly to assure the strength of the QAES, the algebraic and quantum attacks are going to be implemented, and the results are going to be analysed.

## Acknowledgment

The authors would like to thank the anonymous reviewers for their valuable comments and suggestions that improved the presentation of this paper.

## References

1. M. Sumalatha, M. Malal, Design of high speed 128 bits AES algorithm for data encryption, International Journal of current engineering and technology (IJCET), Vol. 20, No. 41, pp 338-343 (2013)
2. A. Kazys, K. Janus, Key-dependent S-box generation in AES block cipher system, Institute of mathematics and informatics, Informatica, Vol. 20, No. 1, pp 23-34 (2012).
3. S. Shaaban, E. Wisam, A. Shadi, Enhancement the security of AES against modern attacks by using variable key block cipher, International Arab journal of e-technology, Vol. 3, No.1, pp 17-26 (2013)
4. H. Rose, J. Hag, Using a cipher key to generate dynamic S-box in AES cipher system, international journal of computer science and security, Vol. 6, No. 1, pp 19-28 (2012)
5. A. Sekar, S. Radhika, K. Anand, Secure communication using 512 bit key, European journal of scientific research, Vol. 52, No.1, pp 61-65(2012)
6. R. Kinga, A. Aline, E. Christian, Generation and testing of random numbers for cryptographic application, Proceeding of Romanian Academy, Vol. 13, No. 4/2012, pp 368-377.
7. J. Daemen and V. Rijmen, The Design of Rijndael: AES -The Advanced Encryption Standard of Information Security and Cryptography, Springer Verlag, Vol. 22, No.5, (2002)
8. D. Warren, W. Smith, “ AES seems weak, linear time secure cryptography “, (online available),
9. William Stallings, Cryptography and Network Security Principles and Practice,pp. 330-337, publishing as Prentice Hall, 5th edition, ISBN 10: 0-13-609704-9,(2012).
10. B. Alex, G. Johann, Cryptanalysis of the Full AES Using GPU-Like Special-Purpose Hardware, Journal Fundamental Informatics - Cryptology in Progress: 10th Central European Conference on Cryptology,Vol. 114, No. 3-4, pp 221-237(2012).
11. D. Bernstein, H. Chen, M. Chen, C. Cheng, C. Hsiao, T. Lange, The billion-mulmod-per-second PC. In SHARCS ’09: Special-Purpose Hardware for Attacking Cryptographic Systems, Lausanne,pp 131–144(2009).
12. S. Hadi, S. Alireza, B. Behnam, A. Mohammadreza, Cryptanalysis of 7-Round AES-128, international journal of computer application, Vol. 10, No. 23, pp 21-29(2013).
13. A. Biryukov, I. Nikolic, Automatic search for related-key differential characteristics in byte-oriented block ciphers: Application to AES, H. Gilbert, editor, Advances in Cryptology EUROCRYPT 2010, Lecture Notes in Computer Science, Springer, Verlag, Vol. 6610, pp 322–344(2010).
14. W. Geiselmann, F. Januszewski, H. K¨opfer, J. Pelzl, R. Steinwandt, A simpler sieving device: Combining ECM and TWIRL, In M. S. Rhee and B. Lee, editors, Information Security and Cryptology — ICISC 2006, Lecture Notes in Computer Science, Vol. 4296, Springer Verlag, pp 118–135(2007).
15. S. Neetu, “Cryptanalysis of Modern Cryptographic Algorithms”, International Journal of Computer Science and Telecommunications, Vol. 1, No. 2 pp 166-169(2010) .
16. J. Julia, M. Ramlan, S. Salasiah, R. Jazrin, Enhancing AES S-box generation based on a round key, International journal of cyber-security and digital forensics, Vol. 1, No. 3, (2012).
17. C. Amit, T. Damodar, “Analysis of AES Algorithm using symmetric cryptography”, International Journal of Computing, Communications and Networking, Vol. 1, No.2, pp 57-62(2012).

18. P. Payal, D. Soni, An invention of quantum cryptography over the classical cryptography for enhancing security, international journal of applied or innovation in engineering and management, Vol. 2, No. 2, pp 243-246(2013).
19. E. Chip, P. David, T. Gregory, Quantum cryptography in practice, international journal of theoretical security, Vol. 3, No. 10, pp 20-32(2003).
20. S. Arias, N. Merabtine, M. Benslama, A new accurate quantum cryptography control error reconciliation (QCCER) with XOR operator in BB84 protocol, ICIC Express Letters, Vol. 2, No. 2, pp 187-192(2008).
21. R. Michael, A revolutionary security technology, ISSA journal, Vol. 3, No. 6, pp 20-28(2012).
22. L. Bo, Z. Baokang, W. Shilling, W. Chongqing, S. Jinshu, Y. Wanrong, W. Fei, Q phone: A quantum security, VOIP phone, SIGCOMM'13, ACM 978-1-4503-2056-6/13/08, pp 477-480(2013).
23. Michel Cukier, Robin Berthier, Susmit Panjwani, Stephanie Tan, A Statistical Analysis of Attack Data to Separate Attacks, International Conference on Dependable Systems and Networks (DSN 2006), 25-28 June 2006, Philadelphia, Pennsylvania, USA, pp. 123-142(2006).
24. Z. Xiao, R. Timothy, K. Pruet, Z. Mian, P. Alberto, Adding the control to arbitrary unknown quantum operations, international journal of nature communications, doi: 10.1038/ncoms1392, pp 1-8(2012).
25. Widmer L., Cerberis: High-Speed Encryption with Quantum Cryptography, published in Advances in Intelligent and Soft Computing book, springer, 2009,PP 31-238.
26. Omer K. Jasim, Safia Abbas, El-Sayed M. El-Horbaty and Abdel-Badeeh M. Salem, "A Comparative Study between Modern Encryption Algorithms based On Cloud Computing Environment", the 8th International Conference for Internet Technology and Secured Transactions (ICITST-2013), UK, Dec.,2013, pp.536-541.

**Authors Biography**

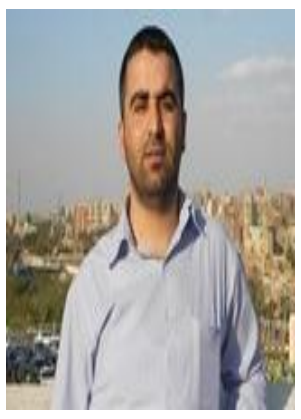

**Omer K. Jism:** He received his M.Sc. (2009) and B.Sc. (2007) in Computer Science from Al-Anbar University, Iraq. He got the first ranking in two above studies with honor grade. His work experience includes 3 years as an academic in Iraq (Alma'arif University College). He worked as Head of the Computer science department, Alma'arif University College (2010-2012) and he is senior member of IEEE. Now he a Ph.D. student in Ain Shams University. Research interesting for omer are, cloud computing, quantum cryptography, cryptography and network security, and artificial intelligent.

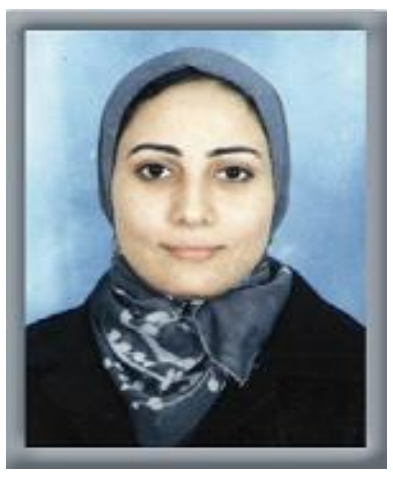

**Dr. Safia Abbas**: She received his Ph.D. (2010) in Computer science from Niigata University, Japan, her M.Sc. (2003) and B.Sc.(1998) in computer and information sciences from Ain Shams University, Egypt. Research interesting for Safia include data mining argumentation, intelligent computing, and artificial intelligent. She has published around 15 papers in refereed journals and conference proceedings in these areas which DBLP and springer indexing. She holds the first place in the international publication with honor from the president of Ain Shams University.

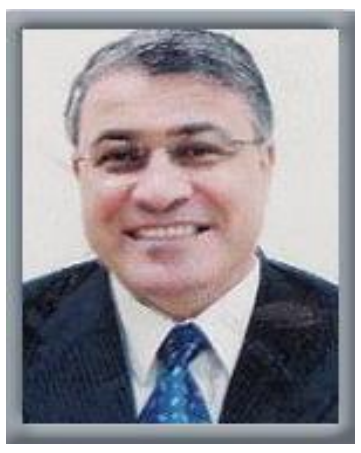

**Professor El-Sayed M. El-Horbaty**: He received his Ph.D. (1985) in Computer science from London University, U.K., his M.Sc. (1978) and B.Sc.(1974) in Mathematics from Ain Shams University, Egypt. His work experience includes 25 years as an academic in Egypt (Ain Shams University), Qatar (Qatar University), and Emirates (Emirates University, Ajman University, and ADU University). He Worked as Deputy Dean of the faculty of IT, Ajman University (2002-2008). He is working as a Vice Dean of the faculty of Computer & Information Sciences, Ain Shams University (2010-2012), and he is working as a head of computer science dept. in the same college (2012-now). Prof. El-Horbaty is current areas of research are parallel algorithms, combinatorial optimization, image processing. His work appeared in journals such as Parallel Computing, International Journal of Computers and Applications (IJCA), Applied Mathematics and Computation, and International Review on Computers and Software. Also he has been involved in more than 26 conferences.

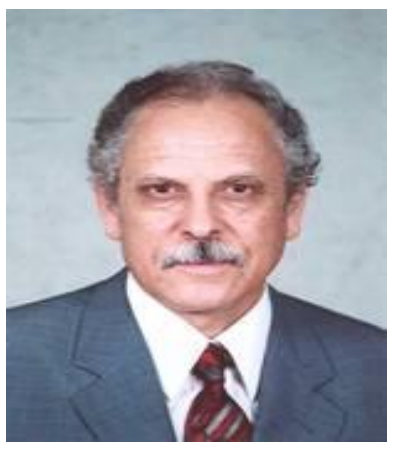

**Prof. Dr. Abdel-Badeeh M Salem**: he is a Professor of Computer Science in the Faculty of Computer and Information Sciences, Ain Shams University, Cairo, Egypt. He has been Professor Emeritus since 2007. Previously he was Director of the Scientific Computing Center at Ain Shams University (1984-1990). His research interests include intelligent computing, expert systems, biomedical informatics, and intelligent e-learning technologies. He has published around 250 papers in refereed journals and conference proceedings in these areas. He has been involved in more than 300 conferences and workshops as an International Program Committee member, organizer and Session Chair. He is author and co-author of 15 Books in English and Arabic Languages. His work appeared in more than 100 journals.